\def\aj{{AJ}}

\def\apj{{ApJ}}
\def\apjs{{ApJS}}

\def\mnras{{MNRAS}}
\def\nat{{Nature}}

\documentclass[cup5b]{caps}
\usepackage{graphicx}
\usepackage{amssymb}
\usepackage{ociwsymp3}   

\HeadText{W. J. Couch, M. M. Colless, and R. De Propris}

\begin{document}

\pagenumbering{arabic}

\author[]{WARRICK J. COUCH$^{1}$, MATTHEW M. COLLESS$^{2}$, and ROBERTO DE PROPRIS$^{2}$ \\
(1) School of Physics, University of New South Wales, Sydney, Australia\\
(2) Research School of Astronomy \& Astrophysics, Australian National 
University, ACT, Australia}

\chapter{Clustering Studies with the 2dF \\ Galaxy Redshift Survey}

\begin{abstract}
The 2dF Galaxy Redshift Survey has now been completed and has mapped the 
three-dimensional distribution, and hence clustering, of galaxies in exquisite 
detail over
an unprecedentedly large ($\sim 10^{8}\,h^{-3}$\,Mpc$^{3}$) volume 
of the local Universe. Here we highlight some of the major results to
come from studies of clustering within the survey: galaxy correlation 
function and power spectrum analyses and the constraints they have placed
on cosmological parameters; the luminosity functions of rich galaxy
clusters, their dependence on global cluster properties and galaxy
type, and how they compare with the field; and the variation of galactic
star formation activity with environment, both within clusters and in
galaxy groups.
\end{abstract}

\section{Introduction}

Given the long and distinguished record the Carnegie Observatories have in 
the exploration of galaxies and determining their distances from us, it is
perhaps fitting that this centennial celebration coincides with the 
emergence of the new generation of large galaxy redshift surveys that are 
now mapping the galaxy distribution over statistically representative 
($\sim 10^{8}\,h^{-3}$\,Mpc$^{3}$) volumes of the local Universe. The 2dF 
Galaxy Redshift Survey (2dFGRS; Colless et al. 2001), carried out on the 
3.9\,m Anglo-Australia Telescope at Siding Spring Observatory, Australia, is 
one such survey. Unlike the Sloan Digital Sky Survey (Nichol 2003),
the 2dFGRS has already been completed, with its final 
observations taken on 11 April 2002. At that point it had obtained spectra 
for 270,000 objects, providing redshifts for 221,496 unique galaxies in
the range $0.0\leq z\leq 0.3$, with a median of $z=0.11$. 

The survey provides an almost complete sampling of the galaxy distribution 
down to an extinction-corrected limit of $b_{J}=19.45$ mag over $\sim 
1800$\,square degrees of sky. This sky coverage is contained within two 
declination strips covering $75^{\circ}\times 10^{\circ}$ and $80^{\circ}\times
15^{\circ}$ in the NGP and SGP regions, respectively, plus 99 ``random''
2-degree fields in 
the SGP. The contiguous coverage of the sky within the two strips, together 
with the almost complete ($\sim 95$\%) sampling of the galaxy distribution, 
has produced the highest fidelity three-dimensional (3D) maps ever seen of the galaxy 
distribution and its large-scale structure. This is clearly revealed in 
the cone diagram shown in Figure 1.1, where the full range of structures
(knots, filaments, voids) are resolved in fine and delicate detail. 

One of the most conspicuous features of the galaxy distribution seen in 
Figure 1.1 is the {\it clustering} of galaxies, and the abundance of dense, 
``knotty'' structures in which rich clusters are embedded. It is the 
quantification of this clustering and the rich wealth of data on clusters 
(and their environs) that the 2dFGRS has provided that is the focus of
this paper. Firstly, we briefly review the key findings to come from the  
precision measurements of galaxy clustering afforded by 2dFGRS, many of 
which have become the ``flagship'' results of the survey. We then describe 
the work that we have been doing on {\it known} rich clusters within the 
survey, using them as sites to study the galaxy luminosity function (LF) 
and galactic star formation in the densest environments, and hence via 
contrast with these same properties measured in lower-density regions
within the survey, to draw conclusions as to their environmental dependence. 
Finally, we briefly mention work in progress on using the survey itself
to generate a new 3D-selected catalog of galaxy groups and clusters using 
automated and objective group-finding algorithms. 

\begin{figure*}[t]
\includegraphics[width=1.00\columnwidth,angle=0,clip]{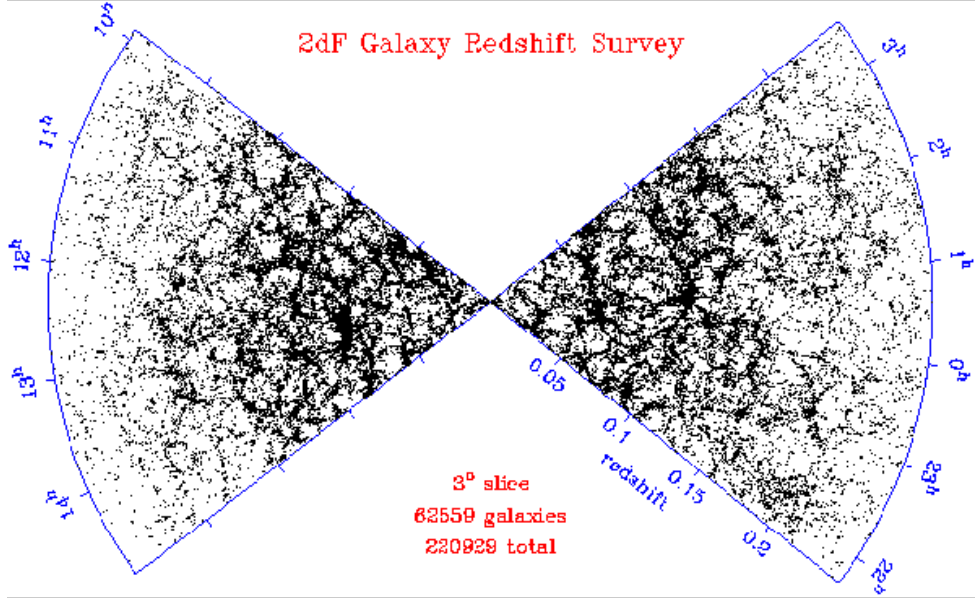}
\vskip 0pt \caption{
A cone diagram based on a 3-degree slice taken through the 
NGP ({\it left}) and SGP ({\it right}) strips of the 2dFGRS.}
\label{}
\end{figure*}

\section{Galaxy Clustering: Key Results}

In quantifying and characterizing the clustering seen in the 3D galaxy
distribution, 2dFGRS has realized significant advances, not just in
the precision of the measurements, but also in extending them to much
larger scales. The key to the latter is the much larger volumes that
are (sparsely) probed by the random fields (cf. the strips on their
own), allowing structure to be measured on scales up to $400\,h^{-1}$\,Mpc.

The clustering over these scales has been measured statistically using both 
two-point correlation function and power spectrum analyses. Being based on 
redshift information, these functions are, of course, derived in terms of the
redshift-space rather than the real-space positions of galaxies. 
Nonetheless, when used jointly and in combination with the matter power 
spectrum provided by cosmic microwave background measurements, they have yielded a number of new
fundamental results on the origin of large-scale structure, the matter 
content and density of the Universe, and galaxy biasing; these can be
summarized as follows (for further details, see the specific references
quoted and also Colless 2003):

\begin{enumerate}
\item Unambiguous detection of coherent collapse on large scales ---
manifested by a flattening of the 2D correlation function in 
the line-of-sight direction at scales of 20--40\,$h^{-1}$\,Mpc --- 
confirming structures grow via gravitational instability (Peacock et al. 
2001).
\item The detection in the power spectrum of ``acoustic'' oscillations 
due to baryon-photon coupling in the early Universe. Derivation from these 
oscillations of the baryon fraction: $\Omega_{\rm b}/\Omega_{\rm m} = 
0.17\pm 0.06$ (Percival et al. 2001).
\item Measurement of $\Omega_{\rm m}$ from both the power spectrum and the 
redshift-space distortions seen in the 2D correlation function: 
$\Omega_{\rm m}=0.30\pm 0.06$. 
\item Through comparison of the 2dFGRS power spectrum and the cosmic microwave background 
power spectrum, the first measurement of the galaxy bias parameter $b^{*} 
= 0.96\pm 0.08$ (Lahav et al. 2002; see also Verde et al. 2002) and its
variation with galaxy luminosity ($b/b^{*}=0.85+0.15L/L^{*}$; Norberg et 
al. 2001) and type (Madgwick et al. 2003).
\item Placement of a stronger limit on the neutrino fraction, 
$\Omega_{\nu}/\Omega_{\rm m} < 0.13$, implying a limit on the mass of all 
neutrino species of $m_{\nu} < 1.8$\,eV (Elgar{\o}y et al. 2002).
\end{enumerate}

\section{Cluster Luminosity Functions}

\begin{figure*}[t]
\includegraphics[width=1.00\columnwidth,angle=0,clip]{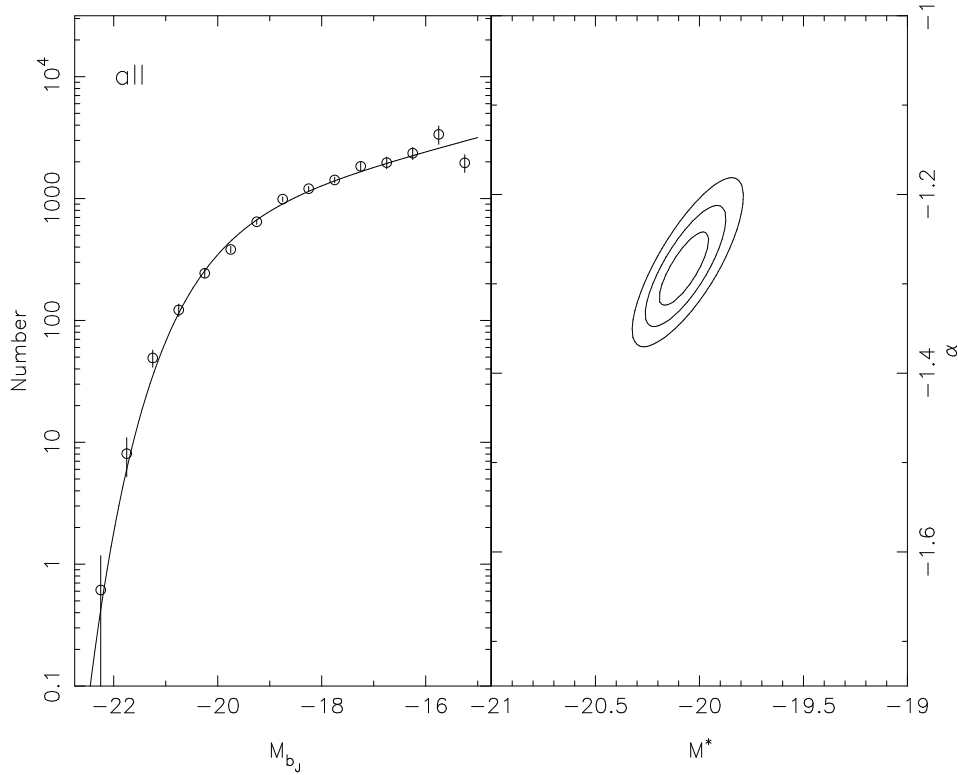}
\vskip 0pt \caption{
{\it Left panel:} The composite LF measured for 60 known rich clusters in the 
2dFGRS. {\it Right panel:} Error contours for the $M^{*}$ and $\alpha$ 
parameter values from a Schechter function fit to this composite LF.}
\label{}
\end{figure*}

The 2dFGRS includes an abundance of rich clusters, with 947 known clusters 
from the Abell (Abell 1958; Abell, Corwin, \& Olowin 1989), APM (Dalton et 
al. 1997) and EDCC (Lumsden et al. 1992) catalogs identified and further
characterized by De~Propris et al. (2002) at a time when 
the survey was just half complete. Of these, 60 have since been used for a 
detailed study of the cluster LF (De~Propris et al. 2003). 
Here the selection was restricted to clusters with $z < 0.11$ in order to 
sample well below the predicted $M^{*}$. A further criterion was that 
clusters must contain at least 40 confirmed  members within the Abell 
radius ($1.5\,h^{-1}$\,Mpc). Cluster membership had been determined 
previously in the De~Propris et al. (2002) study using a ``gapping'' algorithm to isolate 
cluster galaxies in redshift space. The clusters were also chosen to 
ensure a range in velocity dispersion (and hence mass), richness, 
Bautz-Morgan (B-M) type, and structural morphology. 

This sample was used to produce a series of ``composite'' LFs, with clusters 
and their member galaxies being combined to allow meaningful and statistically 
robust comparisons based on both global cluster properties and galaxy type.
As a starting point, a composite LF was derived for the entire cluster
sample containing 4186 members; this is shown in Figure 1.2. As can be seen, 
it provides a very high-quality ``overall'' cluster LF covering 7.5\,magnitudes
in luminosity ($-22<M_{b_{J}}<-16$). It is well fit by a Schechter (1976) 
function with a characteristic magnitude of $M^{*}_{b_{J}}=-20.07\pm 0.07$ 
and a faint-end slope of $\alpha =-1.28\pm 0.03$ (right-hand panel of Fig. 
1.2). 

\begin{figure*}[t]
\includegraphics[width=1.00\columnwidth,angle=0,clip]{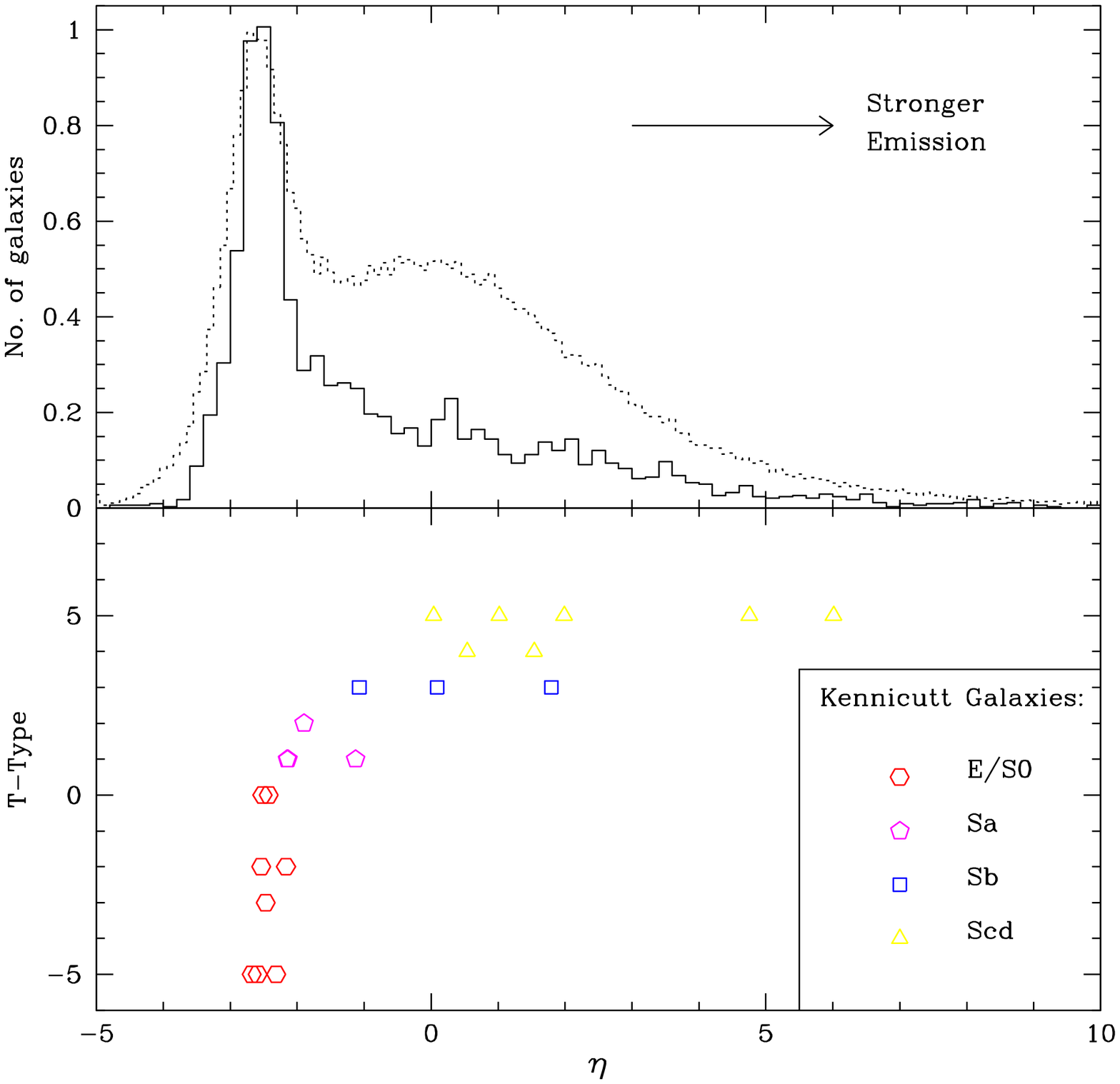}
\vskip 0pt \caption{
{\it Upper panel:} Distribution of different spectral types 
(as measured by the $\eta$ parameter; see text for details) within 
our clusters ({\it solid line}) and in the general field ({\it dotted 
line}). {\it Lower panel:} Relationship between galaxy morphology,  
represented by T type, and $\eta$.}
\label{}
\end{figure*}

A series of composite LFs based on global cluster properties were
also constructed, with the cluster sample divided into subsamples according to 
high ($\sigma_{\upsilon}\ge 800$\,km\,s$^{-1}$)/low 
($\sigma_{\upsilon}<800$\,km\,s$^{-1}$) velocity dispersion, rich/poor, 
``early'' B-M (Types I, I--II, II)/``late'' B-M (Types II--III, III), and with 
substructure/without substructure (see De~Propris et al. 2003 for further 
details). Surprisingly, no statistically
significant LF variation was seen between these different subsamples! The only
conspicuous difference seen was between composite LFs formed for the 
inner (core) and outer regions of clusters, with the former having
many more very bright galaxies than the latter.

Comparison of these composite cluster LFs, in particular the ``overall''
function shown in Figure 1.2, with their 2dFGRS field counterparts is readily
available from the work of Madgwick et al. (2002). A Schechter 
function fit to their composite field LF over the same absolute magnitude 
range yields parameter values of $M^{*}_{b_{J}}=-19.79\pm 0.07$ mag and 
$\alpha =-1.19\pm 0.03$. Hence, taken at face value, it would appear that the 
cluster LF is 0.3\,mag {\it brighter} in $M^{*}_{b_{J}}$ and has a steeper 
faint-end slope ($\alpha_{clus} - \alpha_{fld} =-0.1$). 

\begin{figure*}[t] 
\includegraphics[width=1.00\columnwidth,angle=0,clip]{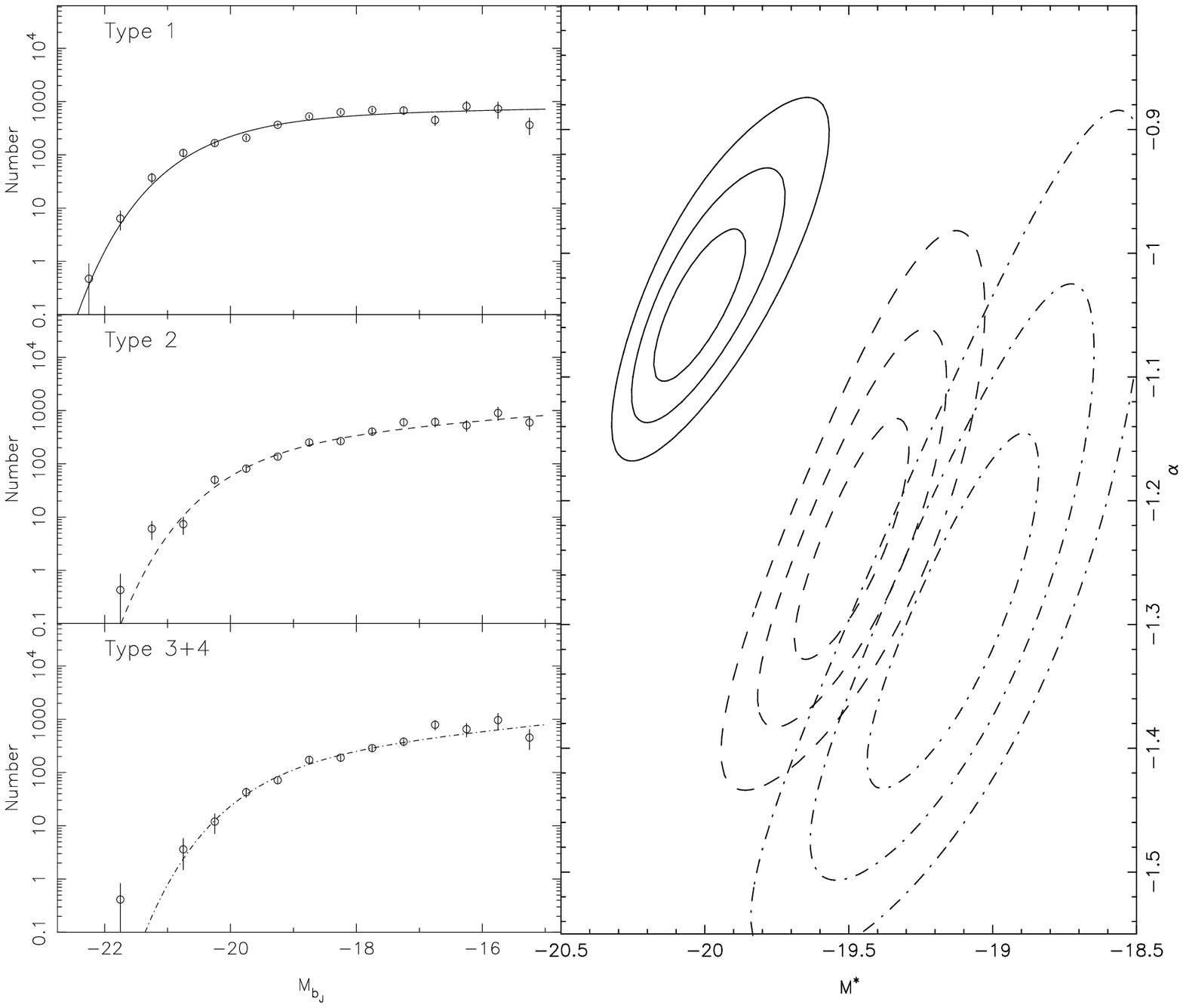}
\vskip 0pt \caption{
{\it Left panel:} LFs for galaxies of different spectral types within 
clusters. {\it Right panel:} The 1, 2, and 3 $\sigma$ error contours for the 
$M^{*}_{b_{J}}$ and $\alpha$ values from the Schechter function fits to these 
LFs.}
\label{}
\end{figure*}

However, clusters contain a very different morphological mix of galaxies 
compared to 
the low-density field (Oemler 1974; Dressler 1980), and before this difference 
in the LFs between the two can be interpreted as evidence for an 
environmental dependence, underlying LF-galaxy type effects (which have
already been seen in the field; Folkes et al. 1999; Madgwick et al. 2002) 
need to be first investigated. This has been done on the basis of galaxy
spectral type, paralleling the approach taken by Madgwick et al. (2002) for
the field. Here, a galaxy's spectrum is typed on the basis of the relative 
strength of its first two principal components (from a principal component
analysis analysis; Folkes et al. 1999), which represent the emission
and absorption components within the spectrum. This is parameterized in
terms of the quantity $\eta$, which is the linear combination of these
two components: $\eta = a\,pc_{1}-pc_{2}$. As might be expected, clusters 
and the field have quite a different mix of galaxies in terms of $\eta$, 
as can be seen in the upper panel of Figure 1.3. In line with the 
morphology-density relation (Dressler 1980) and the relationship between
$\eta$ and morphology (T type) that is seen in the bottom panel of
Figure 1.3, clusters are dominated by galaxies with the lowest $\eta$ values
(absorption-line dominated, no emission), whereas the field contains a
much larger proportion of galaxies with higher ($\eta >0$) values, indicative 
of line emission in what are most likely late-type spirals.  For the purposes 
of deriving LFs for different spectral types, Madgwick et al. (2002)
divided the $\eta$ scale into 4 intervals, with Type 1 galaxies being 
those in the range $-5<\eta \leq -1.3$, Type 2 galaxies $-1.3<\eta \leq 
1.1$, Type 3 galaxies $1.1<\eta \leq 3.4$, and Type 4 galaxies $\eta > 
3.4$.  

Using these same definitions, De~Propris et al. (2003) derived composite LFs 
for the Type 1, 
Type 2, and Types 3+4 galaxies within their clusters; these are plotted
in the left-hand panel of Figure 1.4. Here we see the same general trends 
that are seen in the field (Madgwick et al. 2002) in that earlier spectral types have LFs with 
brighter characteristic magnitudes and flatter faint-end slopes. This can 
be seen quantitatively in the right-hand panel of Figure 1.4. However, 
careful comparison of these LFs with their field counterparts (Fig. 1.5) 
reveals some subtle, but significant, differences, at a level of detail only
a large survey like 2dFGRS can discern. 

\begin{figure*}[t]
\includegraphics[width=1.00\columnwidth,angle=0,clip]{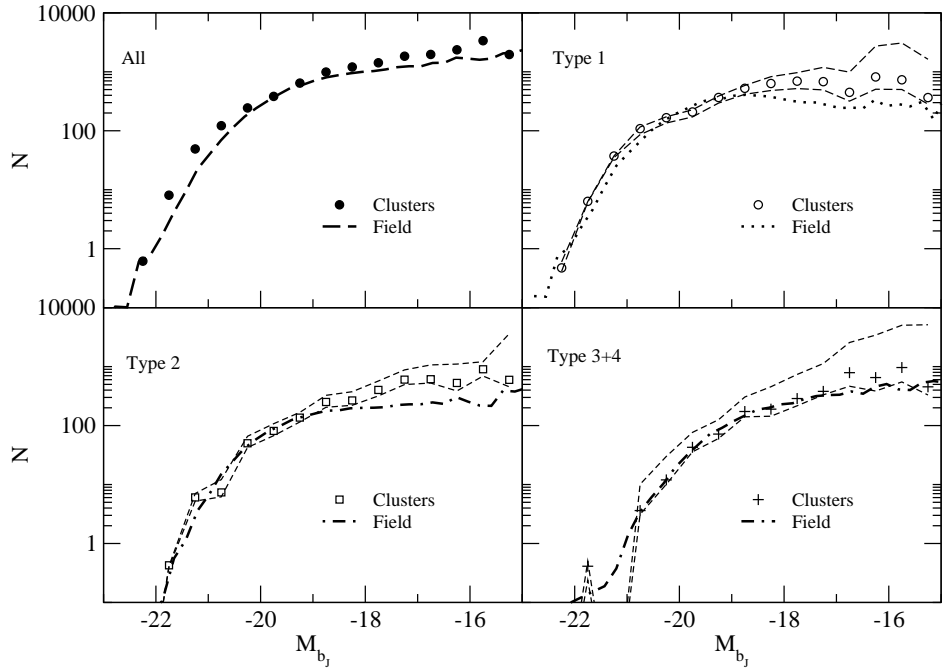}
\vskip 0pt \caption{
Comparison of cluster and field LFs. The {\it points} represent 
the cluster data shown in the previous two figures. The {\it thick}
dashed lines in each panel represent the 2dFGRS field data from Madgwick 
et al. (2002). The {\it thin} dashed lines provide representations of 
the cluster data with no completeness correction ({\it lower}) and 
maximum completeness correction ({\it upper}); see De~Propris et al. (2003) 
for further details.}  
\label{}
\end{figure*}

\begin{figure*}[bt]
\includegraphics[width=0.90\columnwidth,angle=0,clip]{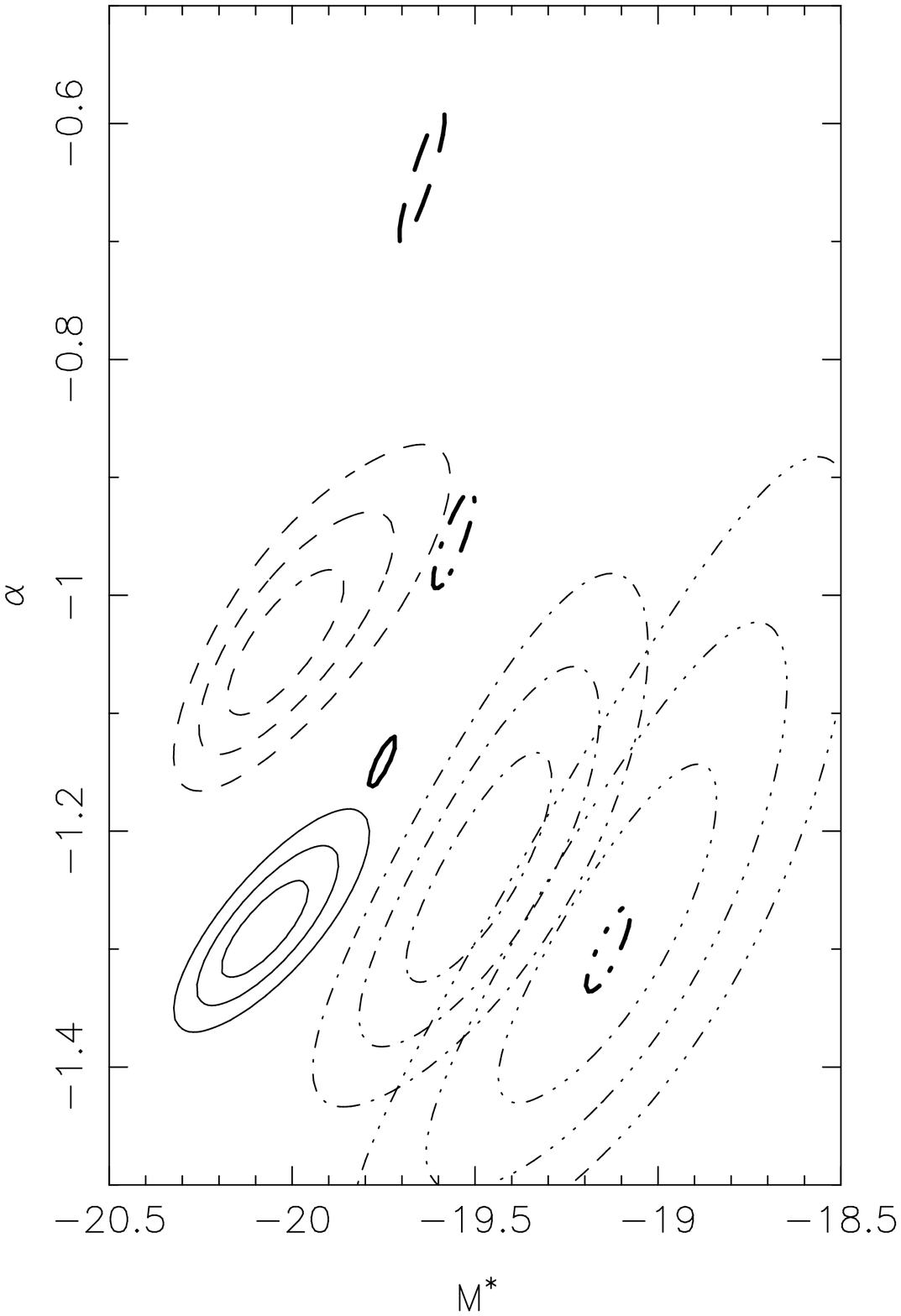}
\vskip 0pt \caption{
Error ellipses for the Schechter function fits to the cluster
({\it thin} lines) and field ({\it thick} lines) LFs. {\it Dashed}
lines: Type 1 galaxies; {\it dot-dashed} lines: Type 2 galaxies;
{\it dotted-dashed} lines: Types $3+4$. Only the 3 $\sigma$ error
ellipse is shown for the field data.}
\label{}
\end{figure*}

These differences are encapsulated in Figure 1.6, which shows the location 
in the $\alpha-M^{*}_{b_{J}}$ plane of the Type 1, Type 2, and Type 3+4
LFs both for clusters and the field, as well as the ``overall'' LFs based
on all types. Clearly the biggest difference between cluster and field 
is for the early types, with the cluster Type 1 LF having a brighter
$M^{*}_{b_{J}}$ (by $\sim 0.5$\,mag) and a steeper faint-end slope 
($\alpha_{cl} - \alpha_{fld} =-0.5$) compared to the field. In contrast, 
the cluster and field LFs for the latest types (Types 3 and 4) are, to
within the errors, indistinguishable. 

The type-dependent differences in the Schechter LF parameters that are
seen between clusters and the field in Figure 1.6 also put the comparison 
of the ``overall'' LFs into perspective. Although there is a significant
difference between the two, it is small in comparison to the differences
seen between the early types (Types 1 and 2). This, together with the
quite different mix of spectral types in clusters and in the field, would
indicate that the similarity in the overall LFs is more by accident than
design, with these mix and spectral type effects conspiring to produce
LFs in the two different environments with characteristic magnitudes and
faint-end slopes that are quite similar.

In order to understand these type-dependent LF differences between clusters
and the field in the context of cluster galaxy and environment-dependent
evolution, an initial attempt has been made to model them using a simple
``closed box'' approach (De~Propris et al. 2003). This involves making the following basic 
assumptions: (1)\,cluster galaxies have evolved from field galaxies contained
within the volume that collapses to become the cluster, (2)\,this evolution
is characterized by a suppression of star formation activity and an
accompanying change in spectral type, and (3)\,the number of galaxies is
conserved (i.e., mergers are neglected). Under these assumptions, the
type-specific field LFs are taken to be the initial LFs within the cluster
volume, with the relative normalizations set by the mix of different types that
are observed. 

This very simple model has some success in that it reproduces the similar
shape but different normalization that is observed for the Type 3+4 LFs
in clusters and in the field. Moreover, it does cause the initial field
LFs of the Type 1 and 2 galaxies to evolve toward the forms they are 
observed to have in clusters. However, at a more detailed level, it fails in
producing too many bright Type 2 cluster galaxies, too few very bright
Type 1 cluster galaxies, and too few faint Type 1 cluster galaxies. 
Obvious refinements to this model (on which we are currently working) are
to allow for luminosity-dependent fading to steepen the faint-end slopes
of the earlier types, and to include mergers to explain the excess of very
bright early types at the expense of their fainter counterparts. 

\section{Star Formation Versus Environment}

The galaxy spectra obtained by the 2dFGRS are of sufficient 
quality not only to determine redshifts but also to measure spectral-line 
indices and derive astrophysical information about the galaxies 
themselves. Moreover, the spectra extend to sufficiently 
red wavelengths ($\lambda\approx 8500$ \AA) to include the redshifted H$\alpha$ 
emission (or absorption) line, thus providing a reliable means of measuring 
the overall star formation rate (SFR) within galaxies (Kennicutt 1992). In this 
section we describe how this has been used to track star formation
activity as a function of environment, firstly from a cluster-centric
point of view, and secondly in galaxy groups.

\subsection{Within Rich Clusters}

To date, the global SFR among galaxy populations has 
generally concentrated on the two extremes of galaxy environment: the 
low-density field and the dense cores of rich clusters. For the latter, 
attention has been very much drawn by the discovery of Butcher \& Oemler 
(1978) that such systems harbored many more star-forming galaxies in the 
past. But galaxies in cluster cores comprise ony a small fraction of
the stellar content of the Universe and may be subject to environmental 
effects that are peculiar to these very high-density environments (e.g., 
ram pressure  stripping, galaxy ``harassment,'' and tidal interactions; 
Dressler 2003). Much more pertinent to the evolution of
the general galaxy population is the environment {\it between} cluster 
cores and the field, spanning 3 orders of magnitude in galaxy density and
which remains largely unchartered in terms of tracking star formation.

\begin{figure*}[t]
\includegraphics[width=1.00\columnwidth,angle=0,clip]{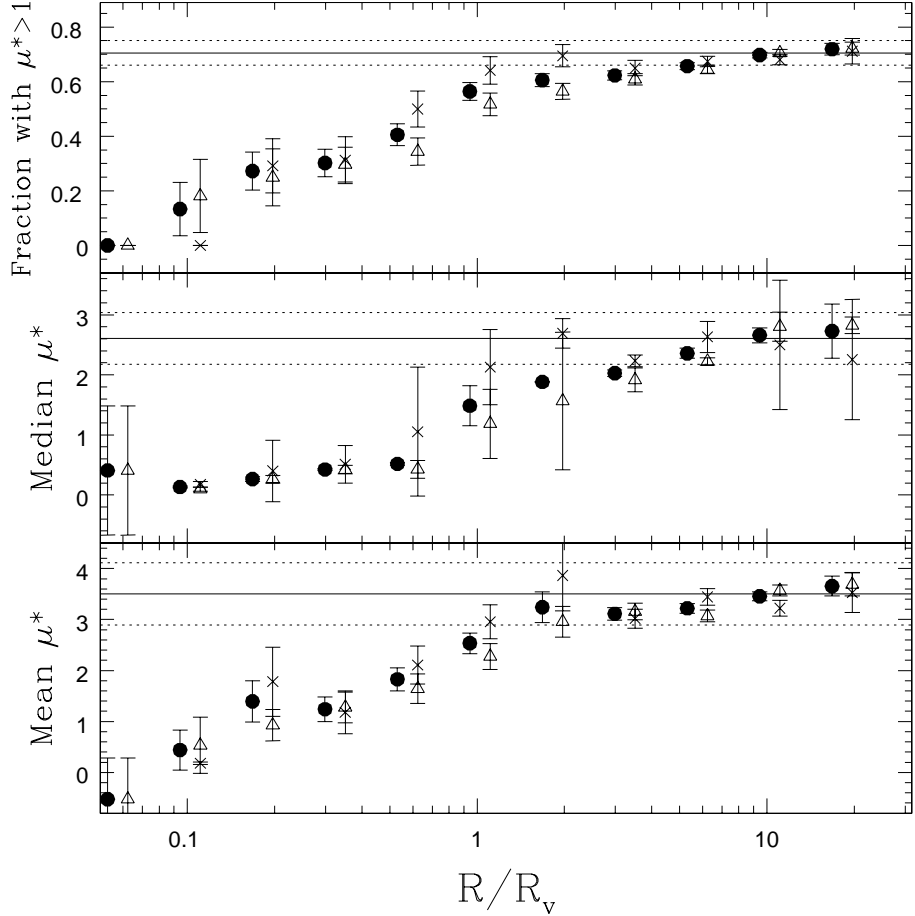}
\vskip 0pt \caption{
Star formation rate as a function of cluster-centric 
radius, derived from our 2dFGRS analysis. {\it Filled} points represent 
the full galaxy sample, while the {\it triangles} and {\it crosses} 
represent the ``high''- and ``low''-velocity dispersion clusters (see text 
for details).  The {\it solid horizontal} lines represent the value of 
each statistic for our field galaxy sample.}
\label{}
\end{figure*}

As a first step toward redressing this situation, we have analyzed 
17 $z\approx 0.05-0.1$ clusters from the De~Propris et al. (2002) study, using 
them to trace the global galactic SFR continuously from their centers out to 
arbitrarily large radii. These clusters were chosen so that roughly half (10) 
had ``high'' velocity dispersions ($\sigma_{\upsilon}>800$\,km\,s$^{-1}$) and 
the other half had ``low'' velocity dispersions ($400\leq\sigma_{\upsilon}\leq 
800$\,km\,s$^{-1}$). All the 2dFGRS galaxies within a projected distance of 
20\,Mpc from the centers of these clusters and within the range 
$0.06<z<0.10$ were selected for analysis. The
restriction in redshift was to limit the effects of aperture bias and poor
sky subtraction. A luminosity cutoff of $M_{b_{J}}=-19$ mag was then applied to
the sample, this being the limit to which 2dFGRS is complete over the adopted 
redshift range. Finally, all galaxies for which there were continuum/line
fitting problems at H$\alpha$ or whose 
EW([N~{\sc ii}] $\lambda$6583)/EW(H$\alpha$) ratios showed evidence of a 
nonstellar component (values $>$0.55) were
excluded from the analysis. This resulted in a final sample of 11,006 galaxies.

The equivalent width of the H$\alpha$ line (be it in absorption or emission)
was measured using an automated Gaussian profile fitting algorithm, which
simultaneously fitted and deblended the H$\alpha$ line from the neighboring
[N~{\sc ii}] $\lambda$6548 \AA\ and [N~{\sc ii}] $\lambda$6583 \AA\ lines. Since only equivalent
width rather than line flux measurements are possible with the 2dFGRS
spectra, we can only infer the SFR, $\mu$, per unit 
luminosity: $\mu/L_{\rm cont}=7.9\times 10^{-42}$\,EW(H$\alpha$), using 
Kennicutt's (1992) SFR-EW(H$\alpha$) relation. We then normalized this
to a characteristic luminosity $L^{*}$: 
$\mu^{*}=\mu/(L_{\rm cont}/L^{*})=0.087$\,EW(H$\alpha$), where $L^{*}$ is
taken to be the knee in the luminosity function in the $r'$ band (near
rest-frame H$\alpha$), as determined by Blanton et al. (2001). 

In Figure 1.7, the mean and median value of $\mu^{*}$ is plotted
as a function of the projected cluster-centric radius. Here the latter has been
normalized by the cluster virial radius, $R_{\upsilon}$, which has been calculated
for each individual cluster and ranges from 1.4 to 2.4\,Mpc. Also plotted in the
top panel of Figure 1.7 is the fraction of galaxies with 
$\mu^{*}>1\,M_{\odot}$\,yr$^{-1}$, which
represents the tail of the distribution, comprised of galaxies that are 
currently forming stars at a high rate relative to their luminosities. 
The solid horizontal line represents the values derived for the ``field''
galaxies within our sample, that is, galaxies that lie within the 20\,Mpc
selection radius, but which, from their redshifts, are
identified as non-members. The bracketing dashed horizontal lines represent
the 1 $\sigma$ standard deviation from field to field, giving some estimate
of the cosmological variance in the field value.

Irrespective of which statistic is used, it is very clear that within 
the clusters $\mu^{*}$ falls significantly below the value of the field, 
the difference being at a maximum at the cluster center and then 
monotonically decreasing with increasing cluster-centric radius. 
{\it Importantly, convergence does not occur until $R>3 R_{\upsilon}$!}
Hence, compared to the field, cluster galaxies differ in their mean
star formation properties as far out as $\sim 6$\,Mpc from their centers. 
Also of note is that these radial trends in $\mu^{*}$ appear to be
insensitive to cluster mass, with there being no discernible difference
between the ``high''- and ``low''-velocity dispersion clusters.

Such a radial analysis, however, is likely to be sub-optimal, since many
of the clusters in our sample are clearly not spherically symmetric and
show substructure. We have therefore analyzed $\mu^{*}$ as a function of 
the local projected galaxy density, $\Sigma$, based on the distance to the tenth
nearest neighbor (Dressler 1980). The relationship between $\mu^{*}$ and
$\Sigma$ is shown in Figure 1.8, using the same three
statistics that we used in the radial analysis. The vertical line shows
the mean value of $\Sigma$ within $R_{\upsilon}$, and once again the horizontal
lines show $\mu^{*}$ and its 1 $\sigma$ variance for the field. 

We see that cluster galaxy star formation is suppressed relative to the field,
the difference being greatest at the highest local densities, and decreasing
monotonically with decreasing density until the two converge at 
$\Sigma\approx 1.5$\,galaxy\,Mpc$^{-2}$, a factor of $\sim 2.5$ times lower
than the mean projected density of the cluster virialized region. Yet
again the trend is no different in the ``high''- and ``low''-velocity 
dispersion clusters, indicating that SFRs depend only
on the local density, regardless of the large-scale structure in which
they are embedded. Further evidence that local density is the key variant
is also provided by a plot based on just those galaxies with $R>2R_{\upsilon}$,
where the same trend observed for the full sample is seen. This would
suggest that a more general view of star formation suppression be taken:
that it will be low relative to the global average in {\it any} region
where the local density exceeds a value of $\sim 1$\,gal\,Mpc$^{-2}$ 
($M_{b}\leq -19$ mag).

\begin{figure*}[t]
\includegraphics[width=1.00\columnwidth,angle=0,clip]{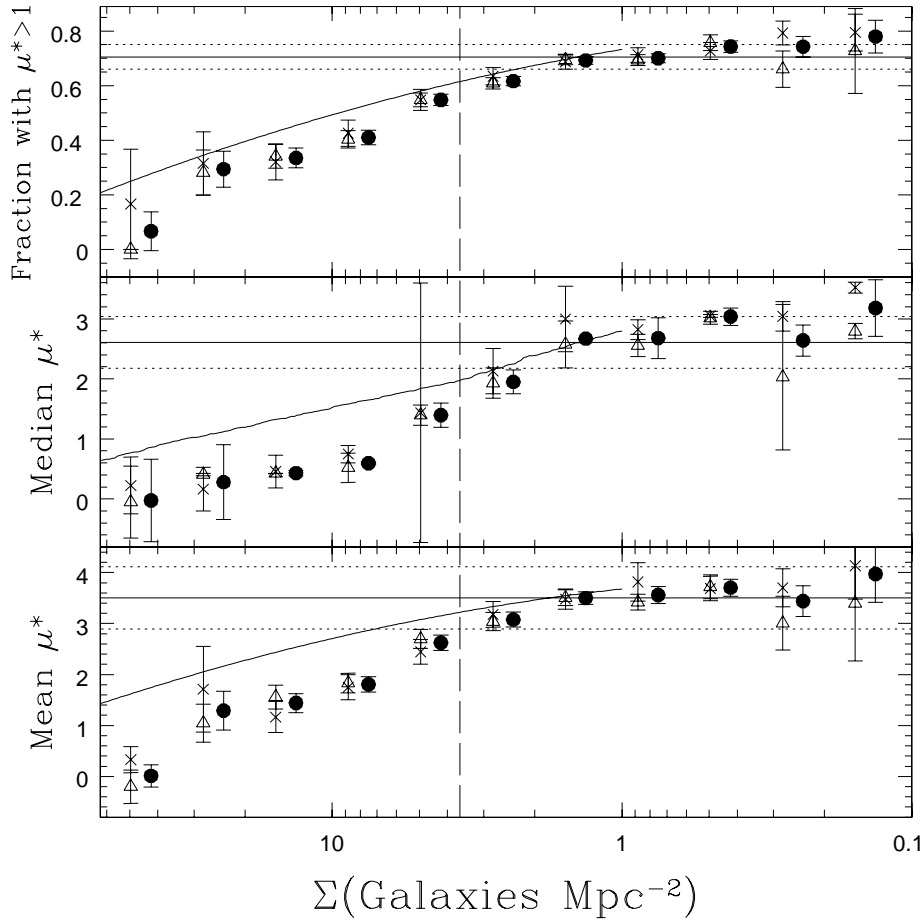}
\vskip 0pt \caption{
Star formation rate as a function of local projected galaxy
density, $\Sigma$. The {\it dashed} vertical line represents the mean
value of $\Sigma$ within $R_{\upsilon}$. The {\it curved} lines indicate the
variation expected as a result of the varying morphological mix.}
\label{}
\end{figure*}

It is well known that galaxy morphology is very strongly correlated with
local galaxy density (Dressler 1980), and hence consideration needs to
be given to what underlying contribution this has to the $\mu^{*}-\Sigma$ 
relationship seen in Figure 1.8. Using a simple model based
on Dressler's morphology-density relation, we have calculated
the expected variation in $\mu^{*}$; this is represented by the 
{\it solid curves} in Figure 1.8. This appears to be
shallower than the observed relation, suggesting that the morphology-density
relation is distinct from the SFR-density relation,  
with additional processes operating at $\Sigma > 1$ galaxy\,Mpc$^{-2}$ that 
drive $\mu^{*}$ significantly below that expected on the basis of the 
changing morphological mix. Indeed, it may well be the case that the
suppression of star formation is the primary transformation that
occurs with environment, with the change in morphological mix being
a secondary effect (Poggianti et al. 1999; Shioya et al. 2002) 

A complete description of this study and a full discussion of its 
implications can be found in Lewis et al. (2002).

\subsection{Within Galaxy Groups}

\begin{figure*}[t]
\includegraphics[width=1.25\columnwidth,angle=0,clip]{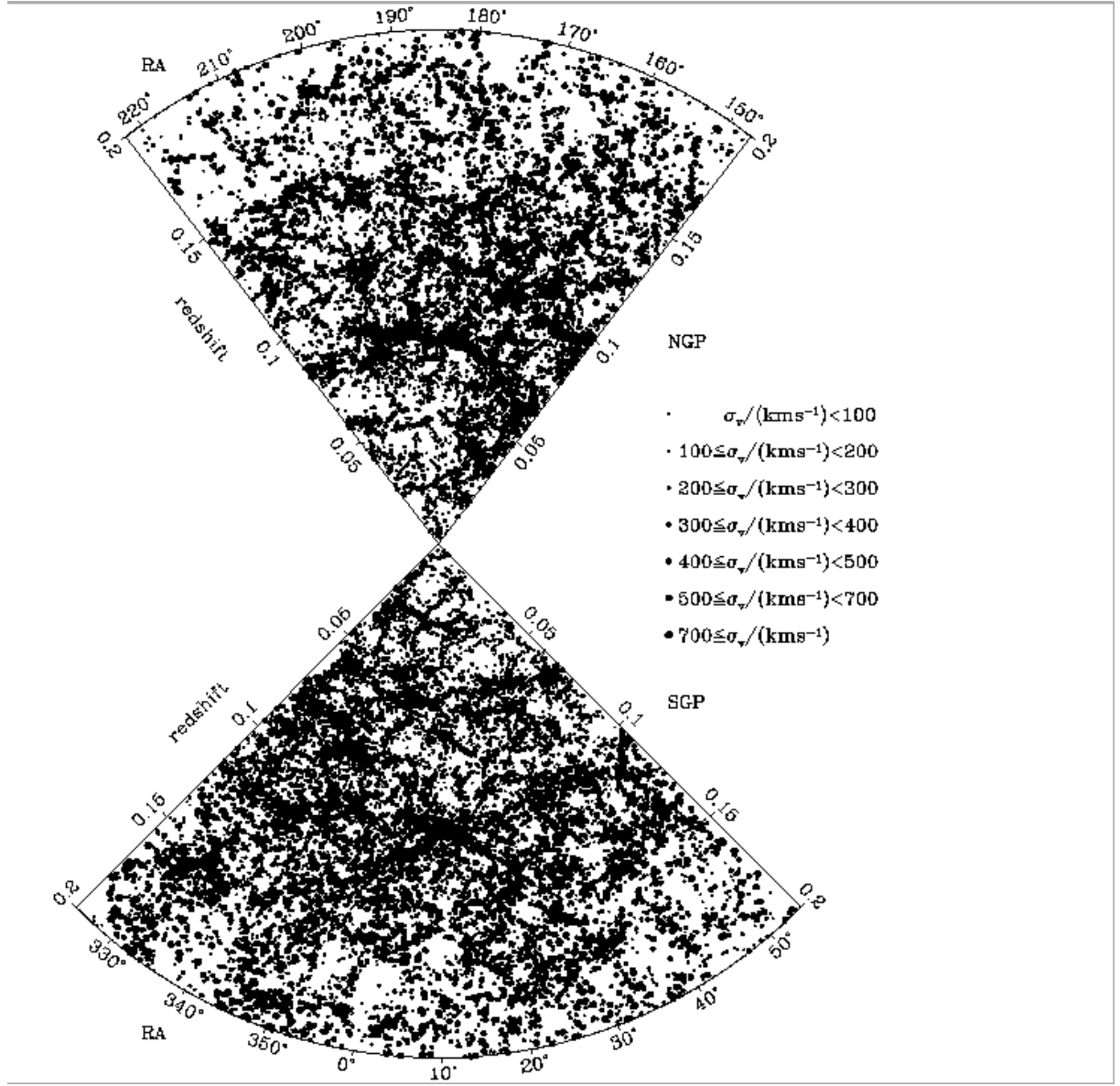}
\vskip 0pt \caption{
A redshift slice showing the distribution of groups and clusters
identified within the 2dFGRS using a 3D ``friends-of-friends'' analysis
(Eke et al. 2003). The estimated velocity dispersion is shown by
the size of the dot.}
\label{}
\end{figure*}

Extension of this analysis to the galaxy group environment has just
recently become possible through the 2dFGRS being subjected to a 
3D ``friends-of-friends'' analysis to identify groups (and clusters)
in both position and redshift space (Eke et al. 2003). This has 
produced a catalog containing $\sim 30,000$ groups with at least 2
members; their distribution within the two 2dFGRS strips is shown
in Figure 1.9. The star formation properties of this sample are
currently being analyzed, and we show here (courtesy of Dr. M. Balogh) 
a couple of very preliminary results that give the first indications of 
environmental trends in these much poorer, lower-velocity dispersion 
analogs to rich clusters.

Once again the H$\alpha$ line has been used as a measure of SFR, with
it being identified and its equivalent width measured in the spectra of
all group members in identical fashion to that described above 
for cluster galaxies. An overall H$\alpha$ equivalent width was then
derived for each group taking the mean value of all its members. In 
Figure 1.10 this mean value is plotted as a function of group velocity
dispersion, being represented by all the individual dots. To assess
whether there is any overall trend, the data have been divided into
equally populated velocity dispersion bins and averaged; the continuous 
line has been drawn through these average values. As a comparison, the mean 
H$\alpha$ equivalent width for ``field'' galaxies --- those galaxies 
identified as not belonging to any group within the friends-of-friends 
analysis --- is shown as the horizontal dashed line. 

\begin{figure*}[t]
\includegraphics[width=1.00\columnwidth,angle=0,clip]{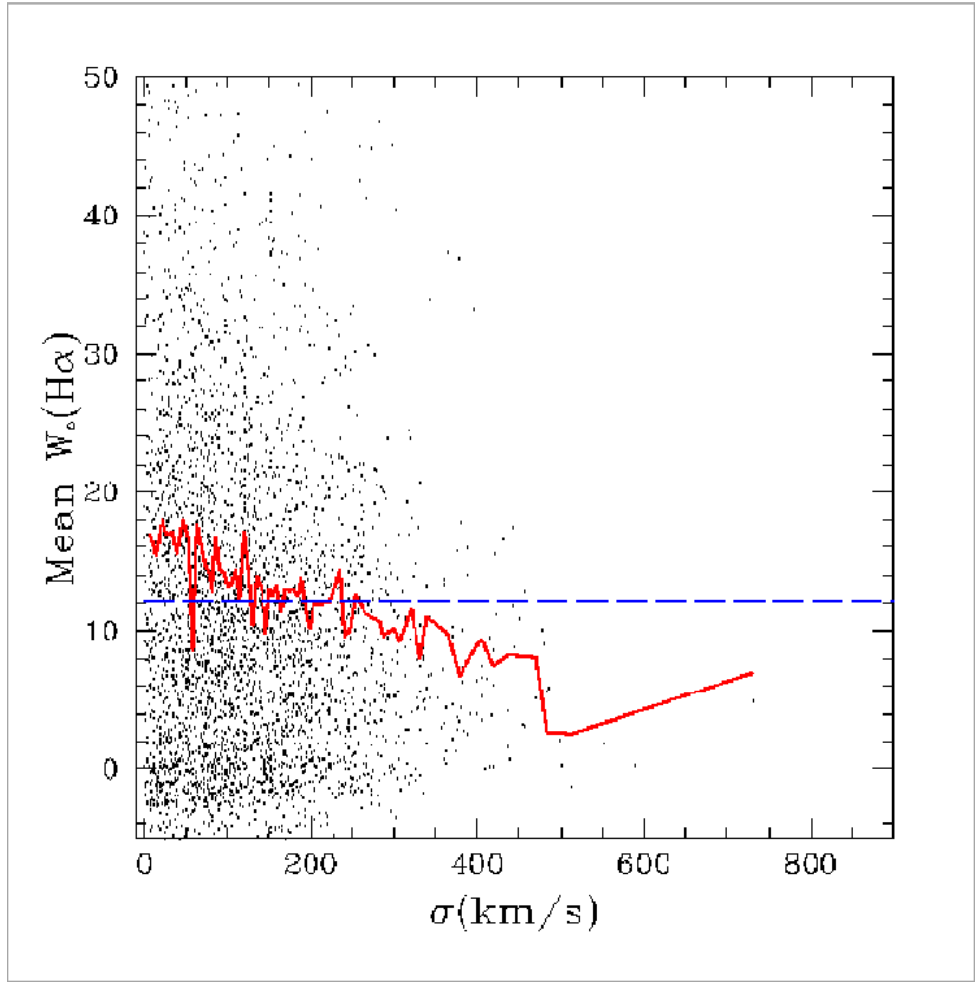}
\vskip 0pt \caption{
The mean equivalent width of H$\alpha$ for groups within
the 2dFGRS versus their velocity dispersion. The {\it dots} represent
individual groups whose values have been determined from their member
galaxies; the {\it solid line} connects the mean values evaluated
within equally populated velocity dispersion bins; the 
{\it dashed} horizontal line indicates the mean value for ``field'' 
galaxies.}
\label{}
\end{figure*}

\begin{figure*}[t]
\includegraphics[width=1.00\columnwidth,angle=0,clip]{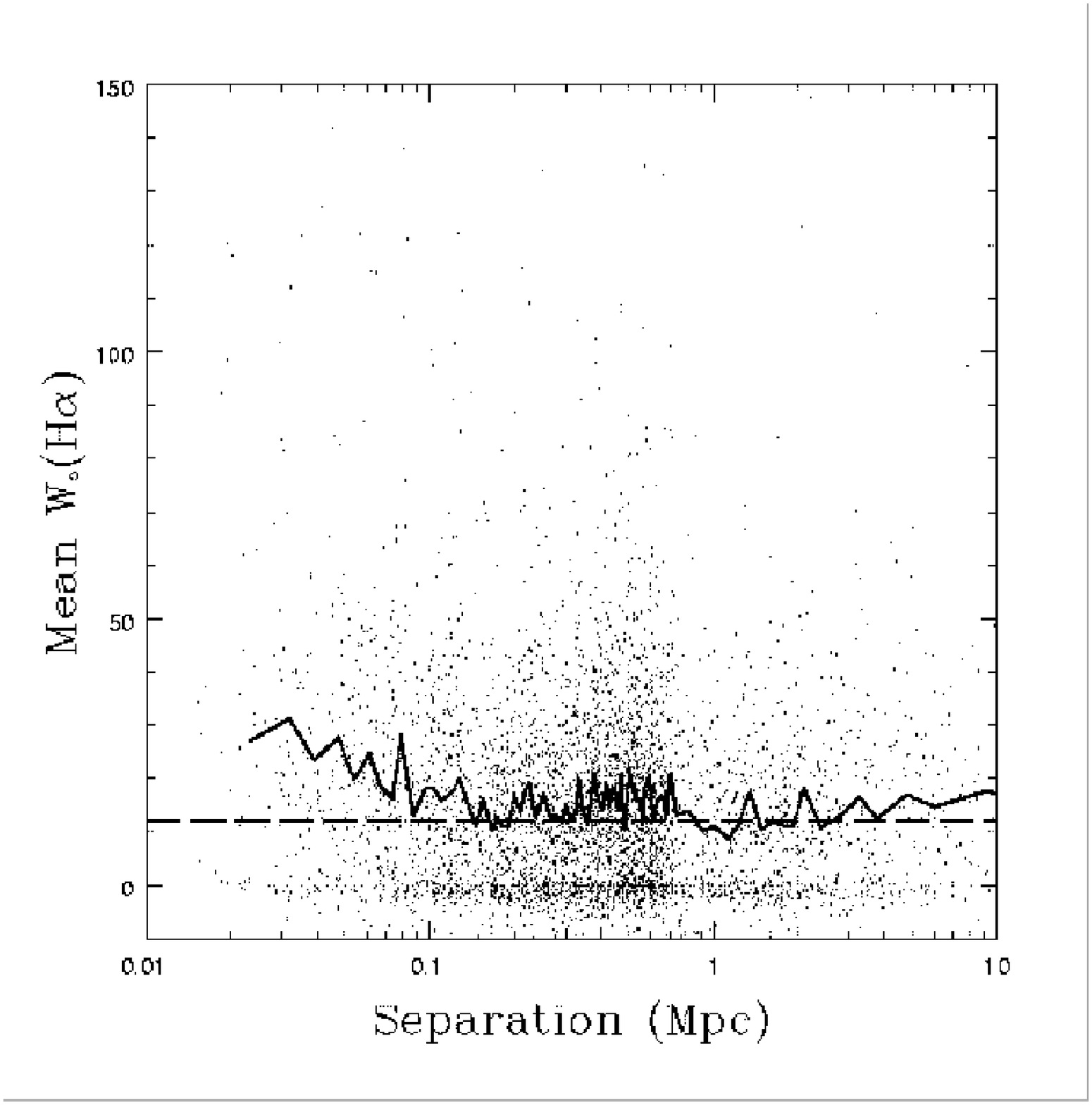}
\vskip 0pt \caption{
The mean equivalent width of H$\alpha$ for binary groups
with velocity dispersions $\sigma < 150$\,km\,s$^{-1}$, plotted as
a function of the separation between the two galaxies. The {\it dots}
represent individual groups; the {\it solid line} connects
the mean values evaluated within equally populated galaxy separation
bins; the {\it dashed} horizontal line indicates the mean H$\alpha$
equivalent width for field galaxies.}
\label{}
\end{figure*}

It can be seen in Figure 1.10 that there is a noisy but significant trend
of H$\alpha$ emission (and hence SFR) becoming increasingly stronger with
lower group velocity dispersion. Intriguingly, however, this trend 
intersects the horizontal ``field'' line at $\sigma\approx 250$\,km\,s$^{-1}$, 
indicating that there are not only groups whose mean EW(H$\alpha$) is below 
that of the field (as might be expected from the SFR--$\Sigma$ relation
observed in the rich cluster study), but also groups whose mean EW(H$\alpha$)
is {\it higher} than the field, hinting that star formation activity
is for some reason enhanced (cf. the field) in these lower-velocity 
dispersion ($\sigma < 150$\,km\,s$^{-1}$) groups. 

The underlying reason for this may well be contained in Figure 1.11, where 
we show the mean EW(H$\alpha$) for just the {\it binary} groups within the 
$\sigma < 150$\,km\,s$^{-1}$ regime, plotted as a function of their 
separation (in Mpc). Yet again, the equal-numbered bin averaging and the
benchmark ``field'' value are represented in the same way as in Figure 1.10. 
We see that for separations in the range 0.1--10\,$h^{-1}$\,Mpc, binary
groups have, to within the fluctuations, the same mean EW(H$\alpha$) as
the field. However, at separations below 0.1\,$h^{-1}$\,Mpc, the mean
EW(H$\alpha$) for binary groups is seen to be significantly higher than the
field. Although this is yet to be fully investigated, it is suggestive 
that the enhanced SFRs seen in the low-velocity dispersion
groups (relative to the field) can be attributed to ``close'' binary systems
where tidal interactions may well be responsible for this activity.

\vspace{0.3cm}
{\bf Acknowledgments}.
These results are presented on behalf of the 2dFGRS team: Ivan Baldry, 
Carlton Baugh, Joss Bland-Hawthorn, Sarah Bridle, Terry Bridges, Russell 
Cannon, Shaun Cole, Matthew Colless, Chris Collins, Warrick Couch, Nicholas 
Cross, Gavin Dalton, Roberto De Propris, Simon Driver, George Efstathiou, 
Richard Ellis, Carlos Frenk, Karl Glazebrook, Edward Hawkins, Carole 
Jackson, Bryn Jones, Ofer Lahav, Ian Lewis, Stuart Lumsden, Steve Maddox, 
Darren Madgwick, Peder Norberg, John Peacock, Will Percival, Bruce 
Peterson, Wil Sutherland, and Keith Taylor. The 2dFGRS was made possible 
through the dedicated efforts of the staff of the Anglo-Australian 
Observatory, both in creating the 2dF instrument and in supporting it on 
the telescope. W.J.C. and R.D.P. acknowledge funding from the Australian
Research Council throughout the course of this work.

\end{document}